\documentclass[11pt,a4paper]{article}
\usepackage{fullpage}
\usepackage{array}
\usepackage{latexsym}
\usepackage{amsfonts}
\usepackage{amssymb}
\usepackage{ipe}
\usepackage{url}
\usepackage[breaklinks,bookmarks]{hyperref}

\long\def\rem#1{}
\newcommand{\empdef}[1]{{#1}}
\newcommand{\refers}[1]{{\rm #1}}

\newcommand{\bra}[1]{\langle #1|}
\newcommand{\ket}[1]{|#1\rangle}
\newcommand{\braket}[2]{\langle #1|#2\rangle}
\newcommand{\norm}[1]{\parallel #1\parallel}
\newcommand{\N}{\mathbb{N}}
\newcommand{\Z}{\mathbb{Z}}
\newcommand{\R}{\mathbb{R}}

\newcommand{\BPP}{\mathrm{BPP}}
\newcommand{\BQP}{\mathrm{BQP}}
\newcommand{\RP}{\mathrm{RP}}
\newcommand{\NC}{\mathrm{NC}}
\newcommand{\AC}{\mathrm{AC}}
\newcommand{\TC}{\mathrm{TC}}
\newcommand{\ACC}{\mathrm{ACC}}
\newcommand{\QNC}{\mathrm{QNC}}
\newcommand{\QAC}{\mathrm{QAC}}
\newcommand{\QACwf}{\mathrm{QAC}_{\mathrm{f}}}
\newcommand{\QTC}{\mathrm{QTC}}
\newcommand{\QTCwf}{\mathrm{QTC}_{\mathrm{f}}}
\newcommand{\QACC}{\mathrm{QACC}}
\newcommand{\QNCwf}{\mathrm{QNC}_{\mathrm{f}}}

\newcommand{\Bc}{\mbox{\rm B-}}
\newcommand{\Rc}{\mbox{\rm R-}}
\newcommand{\mod}{\mathop{\mathrm{mod}}}
\newcommand{\add}{\mathop{\mathrm{add}}}

\newcommand{\Oh}[1]{\mathrm{O}\!\left(#1\right)}
\newcommand{\OO}[1]{\mathrm{O}(#1)}

\newcommand{\Om}[1]{\Omega\!\left(#1\right)}
\newcommand{\Th}[1]{\Theta\!\left(#1\right)}
\newcommand{\poly}{\mathrm{poly}}
\newcommand{\ilog}[1]{\log^{(#1)}}
\newcommand{\logstar}{\log^*}

\newcommand{\Rotate}[2]{\ket{\mu_{\,#1}^{#2}}}
\newcommand{\Rz}[1]{R_{\mathrm z} \left( #1 \right)}

\newcommand{\Phase}[1]{\ket{\psi_{#1}}}
\newcommand{\phase}[1]{\ket{\rho_{#1}}}
\newcommand{\QFTbit}[1]{\phase {x / 2^#1}}

\newenvironment{steps}{%
	\bgroup

	\begin{enumerate}
}{%
	\end{enumerate}
	\egroup
}
\ifx\href\undefined
	\newcommand\email{\begingroup \Url}%
\else
	\newcommand\email[1]{\href{mailto:#1}{\begingroup \Url{#1}}}%
\fi

\newcommand{\floatfigure}[3][tb]{
	\begin{figure}[#1]
	\[
	{\def\IPEfile{pict/#2.ipe}\begingroup
  \catcode`\%=9\catcode`\!=0\catcode`\-=11\input{\IPEfile}}
	\]
	\caption{#3}
	\label{fig:#2}
	\end{figure}
}
\newcommand{\floatfigures}[4][tb]{
	\begin{figure}[#1]
	\[
	#4
	\]
	\caption{#3}
	\label{fig:#2}
	\end{figure}
}

\IpeScale{100}

\newtheorem{definition}{Definition}
\newtheorem{theorem}{Theorem}
\newtheorem{lemma}[theorem]{Lemma}

\newenvironment{remark}[1][Remark.]{
	\par\medskip
	\noindent {\em #1}
}{
	\par\medskip
}
\newenvironment{example}[1][Example.]{
	\par\medskip
	\noindent {\em #1}
}{
	\par\medskip
}

\newenvironment{proof}[1][Proof.]{
	\par
	\noindent \textbf{#1}
}{
	\unskip
	\nobreak\hfill\penalty50\hskip3pt\hbox{}\nobreak\hfill
	\hbox{$\Box$}\par\bigskip
}

\begin{document}

\title{Quantum Fan-out is Powerful}

\author{%
  Peter H{\o}yer%
    \thanks{Supported by Canada's NSERC and the Canadian Institute for
    Advanced Research (CIAR).}
  \\
  Dept.{} of Comp.~Sci., Univ.{} of Calgary\\
  AB, Canada\\
  \email{hoyer@cpsc.ucalgary.ca}
\and
  Robert {\v S}palek%
    \thanks{Work conducted in part while at Vrije Universiteit, Amsterdam.
    Partially supported by EU fifth framework project QAIP, IST-1999-11234
    and RESQ, IST-2001-37559.}
  \\
  Centrum voor Wiskunde en Informatica\\
  Amsterdam, The Netherlands\\
  \email{sr@cwi.nl}
}
\date{}
\maketitle

\begin{abstract}
We demonstrate that the unbounded fan-out gate is very powerful.
Constant-depth polynomial-size quantum circuits with bounded fan-in and
unbounded fan-out over a fixed basis (denoted by $\QNCwf^0$) can approximate
with polynomially small error the following gates: parity, mod[q], And,
Or, majority, threshold[t], exact[t], and Counting.  Classically, we need
logarithmic depth even if we can use unbounded fan-in gates.  If we allow
arbitrary one-qubit gates instead of a fixed basis, then these circuits can
also be made exact in log-star depth.  Sorting, arithmetic operations, phase
estimation, and the quantum Fourier transform with arbitrary moduli can also
be approximated in constant depth.  
\end{abstract}

\section{Introduction}

In this paper, we study the power of shallow quantum circuits.  Long
quantum computations encounter various problems with decoherence, hence
we want to speed them up as much as possible.  We can exploit the following two types
of parallelism:
\begin{steps}
\item Gates on different qubits can be applied at the same time.
\item {\em Commuting} gates can be applied to the {\em same} qubits at the same
time.
\end{steps}

The first approach is just the classical parallel computation.  The second
approach only makes sense when the gates applied on the same qubits commute,
i.e.\ $AB = BA$, otherwise the outcome would be ambiguous.  Being able to do
this is a strong assumption, however there are models of quantum computers, in
which it is physically feasible: ion-trap computers~\cite{cz:qc-ion} and
bulk-spin resonance (NMR)~\cite{gh:qc-nmr}.  The basic idea is that if two
quantum gates commute, so do their Hamiltonians and therefore we can apply
their joint operation by performing both evolutions at the same time.
This type of research started after the M\o lmer--S\o rensen
paper~\cite{ms:ions}.
Recently, a Hamiltonian implementing the fan-out gate (which is crucial for
all our simulations) has been proposed by Fenner~\cite{fenner:fanout}.  

In our paper, we investigate how much the power of quantum computation
would increase if we allow such commuting gates.  The computation in the stronger
model must be efficient, therefore we do not require the ability to perform
{\em any} set of commuting gates.  This is in accordance with standard quantum
computation, where we also allow only some gates.  We choose a representative,
the so-called {\em unbounded fan-out gate}, which is a sequence of controlled-not
gates sharing one control qubit.  We call it fan-out, because if all target
qubits are zero, then the gate copies the {\em classical} source bit into $n$
copies.  We show that fan-out is in some sense universal for all sets of
commuting gates.  In particular, the joint operation of any set of
commuting gates (that can be easily diagonalised) can be simulated by a
constant-depth quantum circuit using just one-qubit and fan-out gates.  To
achieve this, we generalise the parallelisation method of
\cite{moore:parallel-qc,ghmp:qacc} and adapt it to the constant-depth setting.  

We state our results in terms of circuit complexity classes.  Classically, the
main classes computed by constant-depth, polynomial-size circuits are:
\begin{quote}
   $\NC^0$ with Not and bounded fan-in gates: And, Or,
\\ $\AC^0$ with Not and unbounded fan-in gates: And, Or,
\\ $\TC^0$ with Not and unbounded fan-in gates: And, Or, threshold[t]
for all $t$,
\\ $\AC^0[q]$ with Not and unbounded fan-in gates: And, Or, mod[q],
\\ $\ACC^0 = \bigcup_q \AC^0[q]$.
\end{quote}
The zero in the exponent means constant depth, in general $\NC^k$ means
$(\log^k n)$-depth circuits.  Several separations between these classes are
known.  Razborov~\cite{razborov:circsize} proved that $\TC^0$ is strictly more
powerful than $\ACC^0$.  Using algebraic methods,
Smolensky~\cite{smolensky:algmethods} proved that $\AC^0[q] \neq \AC^0[q']$,
where $q, q'$ are powers of distinct primes.  In other words, threshold gates cannot be
simulated by constant-depth circuits with unbounded fan-in Or gates, and
mod[q] gates do not simulate each other.

The main quantum circuit classes corresponding to the classical classes are
$\QNC^0$, $\QAC^0$, $\QTC^0$, and $\QACC^0$.  We use subscript `f' to indicate
circuits where we allow the fan-out gate (e.g.\ $\QNCwf^0$).  Classically,
fan-out (copying the result of one gate into inputs of other gates) is taken
for granted.  Surprisingly, in contrast to the classical case, some of the
quantum circuit classes are the same.  Moore~\cite{moore:fanout} proved that parity
is equivalent to fan-out, i.e.\ $\QACwf^0 = \QAC^0[2]$.  Green et
al.~\cite{ghmp:qacc} proved that allowing mod[q] gates with different moduli
always leads to the same quantum classes, i.e.\ $\QACC^0 = \QAC^0[q]$ for every
integer $q \ge 2$.

In this paper, we extend these results and show that even exact[t] gates
(which output 1 if the input is of Hamming weight $t$, and 0 otherwise) can be
approximated with polynomially small error by fan-out and single qubit gates
in constant depth.  Our simulations have polynomially small error.  Since
exact[t] gates can simulate And, Or, threshold[t], and mod[q] gates, we
conclude that the bounded-error versions of the following classes are equal:
$\Bc \QNCwf^0 = \Bc \QACwf^0 = \Bc \QTCwf^0$.  The exact[t] gate can be
approximated in constant depth thanks to the parallelisation method.  However,
the simulation is not so straightforward as for mod[q] in~\cite{ghmp:qacc} and
it works only with high probability.

We then introduce a so-called Or-reduction that converts $n$ input
bits $x$ into $\log n$ output bits $y$ and preserves the Or function, i.e.\
$x$ is nonzero if and only $y$ is.
We show how to implement it exactly in constant depth and use it to
achieve exact computation of Or and exact[t] in log-star depth.
(Circuits of log-star depth are defined in Section~\ref{sec:log-star}.)
We also apply the Or-reduction to decrease the size of most of our circuits.

Our results concerning the threshold[t] gate have several interesting
implications.  Siu et al.~\cite{sbkh:division-circ} proved that sorting and
integer arithmetic (addition and multiplication of $n$ integers, and division
with remainder) are computable by constant-depth threshold circuits.  It follows
that all of them can be approximated in $\Bc \QNCwf^0$.

The last contribution of our paper concerns the quantum Fourier Transform
(QFT).  Cleve and Watrous~\cite{cw:qft} published an elegant log-depth quantum
circuit that approximates the QFT.  By optimising their methods to use the
fan-out gate, we can approximate the QFT in constant depth with polynomially
small error.  First, we develop a circuit for the QFT with respect to a power-of-2
modulus, and then, using a technique of~\cite{hh:qft}, we show that the QFT with
respect to arbitrary moduli can be approximated too.  Hence the QFT is in $\Bc
\QNCwf^0$.  The QFT has many applications, one of which is the
phase estimation of an unknown quantum state.

Shor's original algorithm for factoring~\cite{shor:factoring} uses the QFT and
the modular exponentiation.  Cleve and Watrous~\cite{cw:qft} have shown that it
can be adapted to use modular multiplication of $n$ integers.  Since we
prove that both the QFT and arithmetic operations are in $\Bc \QNCwf^0$, polynomial-time
bounded-error algorithms with oracle $\Bc \QNCwf^0$ can factorise numbers and
compute discrete logarithms.  We can make the following conclusions:
First, if $\Bc \QNCwf^0$ can be simulated by a $\BPP$ machine, then
factoring can be done in polynomial time by bounded-error Turing
machines.  Second, since it unlikely that $\BQP = \Bc \QNCwf^0$,
factoring and discrete logarithms are likely not the hardest things quantum
computers can do.

\section{Quantum circuits with unbounded fan-out}

{\em Quantum circuits} resemble classical reversible circuits.  A
quantum circuit is a sequence of quantum gates ordered into {\em
layers}.  The gates are consecutively applied in accordance with the order
of the layers.  Gates in one layer can be applied in parallel.  The size
of a gate is the number of affected qubits.  The {\em
depth} of a circuit is the number of layers and the {\em size} is the
total size of all its gates.  A circuit can solve problems of a fixed input size, so we
define {\em families} of circuits containing one circuit for every input
size.  We consider only {\em uniform} families, whose description can be
generated by a log-space Turing machine.

A {\em quantum gate} is a unitary operator applied to some subset of
qubits.  We usually use gates from a fixed {\em universal
basis} (Hadamard gate, rotation by an irrational multiple of $\pi$, and
the controlled-not gate) that can approximate any quantum gate with good
precision~\cite{adh:qcomputability}.
The qubits are divided into 2 groups: {\em Input/output} qubits
contain the description of the input at the beginning and they are
measured in the computational basis at the end.  {\em Ancilla qubits}
are initialised to $\ket{0}$ at the beginning and the circuits usually
clean them at the end, so that the output qubits are in a pure state and
the ancillas may be reused.

Since unitary evolution is reversible, every operation can be undone.
Running the computation backward is called {\em uncomputation} and is
often used for cleaning ancilla qubits.

\subsection{Definition of quantum gates}

Quantum circuits cannot use a naive quantum fan-out gate mapping every
quantum superposition $\ket{\phi}\ket{0}\dots \ket{0}$ to
$\ket{\phi} \dots \ket{\phi}$ due to the no-cloning theorem
\cite{wz:noclone}.  Such a gate is not linear, let alone unitary.
Instead, our fan-out gate copies only classical bits and the effect on
superpositions is determined by linearity.  It acts as a
controlled-not-\dots-not gate, i.e.\ it is an unbounded sequence of
controlled-not gates sharing one control qubit.  Parity is a natural
counterpart of fan-out.  It is an unbounded sequence of controlled-not
gates sharing one target qubit.

\begin{definition}
The \empdef{fan-out gate} maps
\(
\ket{y_1} \dots \ket{y_n} \ket{x} \to
\ket{y_1 \oplus x} \dots \ket{y_n \oplus x} \ket{x},
\)
where $x \oplus y = (x + y) \mod 2$.
The \empdef{parity gate} maps
\(
\ket{x_1} \dots \ket{x_n} \ket{y} \to
\ket{x_1} \dots \ket{x_n} \ket{y \oplus (x_1 \oplus \dots \oplus x_n)}.
\)
\end{definition}

\begin{example}
As used in~\cite{moore:fanout}, parity and fan-out can simulate each other in
constant depth.  The Hadamard gate is
\(
H = {1 \over \sqrt 2} \left( \begin{array}{r r}
	1 & 1 \\
	1 & -1 \\
	\end{array} \right)
\)
and it holds that $H^2 = I$.
If a controlled-not gate is preceded and succeeded by Hadamard gates on
both qubits, it just turns around.  Since parity is a sequence of
controlled-not gates, we can turn around all of them in parallel.
The circuit is shown in Figure~\ref{fig:qncwf-parity2}.
\end{example}
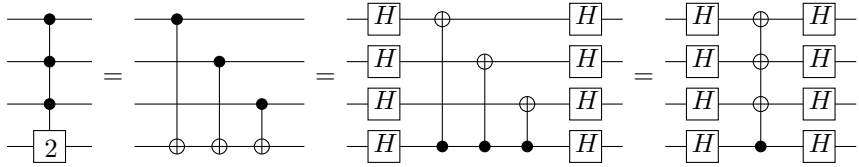
\floatfigure{qncwf-parity2}{Equivalence of parity and fan-out}

\goodbreak
In this paper, we investigate the circuit complexity of, among others, these gates:

\begin{definition}
\label{def:gates}
Let $x = x_1 \dots x_n$ and let $|x|$ denote the Hamming weight of $x$.
The following $(n+1)$-qubit gates map
$\ket{x} \ket{y} \to \ket{x} \ket{y \oplus g(x)}$, where $g(x) = 1$ iff
{\rm \[\vbox{
\halign{# && \strut \hfil $#$: & # \hfil \qquad \cr
& |x| > 0 & Or,
& |x| = n & And (Toffoli),
& |x| \ge \frac n2 & majority,
\cr
& |x| \mod q = 0 & mod[q],
& |x| \ge t & threshold[t],
& |x| = t & exact[t],
\cr
}
}\]}
A \empdef{counting gate} is any gate that maps $\ket{x} \ket{0^m} \to
\ket{x} \ket{\, |x|\, }$ for $m = \lceil \log (n+1) \rceil$.
\end{definition}

\subsection{Quantum circuit classes}

\begin{definition}
\label{def:QNCwf}
$\QNCwf(d(n))$ contains operators computed exactly (i.e.\ without error) by
uniform families of quantum circuits with fan-out of depth $\Oh{d(n)}$,
polynomial size, and over a fixed basis.
$\QNCwf^k = \QNCwf(\log^k n)$.
$\Rc \QNCwf^k$ contains operators approximated with
one-sided, and $\Bc \QNCwf^k$ with two-sided, polynomially small error.
\end{definition}

\begin{remark}
The circuits below are over a fixed universal basis, unless
explicitly mentioned otherwise.  Some of our circuits need arbitrary
one-qubit gates to be exact.  For simplicity, we sometimes include several
fixed-size gates (e.g.\ the binary Or gate and controlled one-qubit
gates) in our set of basis gates.  This inclusion does not influence the
asymptotic depth of our circuits, since every $s$-qubit
quantum gate can be decomposed into a sequence of one-qubit and
controlled-not gates of length $\Oh{s^3 4^s}$ \cite{bbc:gates}.

For every one-qubit gate $U$, there exist one-qubit gates $A, B, C$ and
a rotation $P = \Rz{\alpha}$ such that the controlled gate $U$ is computed
by the constant-depth circuit shown in Figure~\ref{fig:qncwf-control}
\cite[Lemma~5.1]{bbc:gates}.
If a qubit controls more one-qubit gates, then we can still use this
method in constant depth.  We just replace the controlled-not gate by
the fan-out gate and the rotations $P$ are multiplied.
\end{remark}
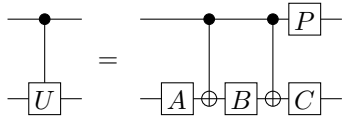
\floatfigure{qncwf-control}{Implementing an arbitrary controlled one qubit
gate}

\section{Parallelisation method}

In this section, we describe a general parallelisation method for achieving
very shallow circuits.  We then apply it to the rotation by Hamming
weight and the rotation by value, and show how to compute them in constant
depth.

\subsection{General method}

The unbounded fan-out gate is universal for commuting gates in the
following sense:  Using fan-out, gates can be applied to the same qubits
at the same time whenever (1) they commute, (2) we know the basis in
which they all are diagonal, and (3) we can efficiently change into the
basis.  The method reduces the depth, but may in general require the use of
ancilla qubits.

\begin{lemma}
\label{lem:commuting}
\refers{\cite[Theorem~1.3.19]{hj:matrix-book}}
For every set of pairwise commuting unitary gates, there exists an
orthogonal basis in which all the gates are diagonal.
\end{lemma}

\begin{theorem}
\label{th:parallel}
\refers{\cite{moore:parallel-qc,ghmp:qacc}}
\newcommand{\dep}[1]{\mathrm{depth}(#1)}
\newcommand{\size}[1]{\mathrm{size}(#1)}
Let $\{ U_i \}_{i=1}^n$ be pairwise commuting gates on $k$ qubits.  Gate
$U_i$ is controlled by qubit $\ket{x_i}$.  Let $T$ be a gate changing the
basis according to Lemma~\ref{lem:commuting}.  There exists a quantum
circuit with fan-out computing $U = \prod_{i=1}^n U_i^{x_i}$ having
depth $\max_{i=1}^n \dep{U_i} + 4\cdot \dep{T} + 2$,
size $\sum_{i=1}^n \size{U_i} + (2n + 2) \cdot \size{T} + 2 n$, 
and using $(n-1)k$ ancillas.
\end{theorem}

\begin{proof}
Consider a circuit that applies all $U_i$ sequentially.  Put $T
T^\dagger = I$ between $U_i$ and $U_{i+1}$.  The circuit is shown in
Figure~\ref{fig:comm-serial+subst}.  Take $V_i = T^\dagger U_i
T$ as new gates.  They are diagonal in the computational basis, hence
they just impose some phase shifts.
Multiple phase shifts on entangled states multiply, so can be applied in
parallel.  We use fan-out gates twice: first to create $n$ entangled
copies of target qubits and then to destroy the entanglement.  The final
circuit with the desired parameters is shown in
Figure~\ref{fig:comm-parallel}.
\end{proof}
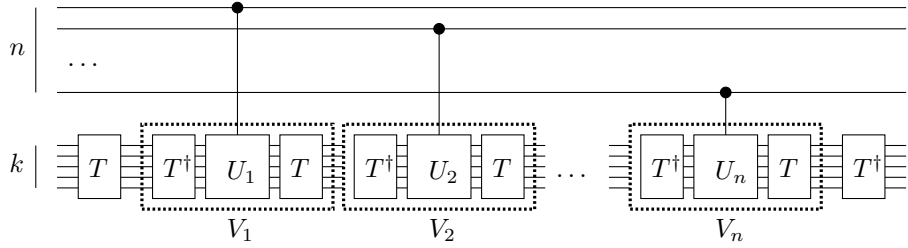
\floatfigure{comm-serial+subst}{A serial circuit with interpolated basis changes}
\floatfigure{comm-parallel}{A parallelised circuit performing
$U = T^\dagger (\prod_{i=1}^n V_i^{x_i}) T = \prod_{i=1}^n U_i^{x_i}$}

\begin{example}
As used in \refers{\cite{moore:fanout}}, it is simple to prove that mod[q] $\in
\QNCwf^0$.  Each input qubit controls one increment modulo $q$ on a
counter initialised to $0$.  At the end, we obtain $|x| \mod q$.
The modular increments commute and thus can be parallelised.  Since $q$
is fixed, changing the basis and the increment can both be done in
constant depth.
\end{example}

\subsection{Rotation by Hamming weight and value}

In this paper, we often use a {\em rotation by Hamming weight} $\Rz{\varphi
|x|}$ and a {\em rotation by value} $\Rz{\varphi x}$, where $\Rz{\alpha}$ is
one-qubit rotation around the $z$-axis by angle $\alpha$: $\displaystyle
\Rz{\alpha} = \ket0 \bra0 + e^{i \alpha} \ket1 \bra1$.  They can both be
computed in constant depth.

\begin{lemma}
\label{lem:rotation}
For every angle $\varphi$, there exist constant-depth, linear-size
quantum circuits with fan-out computing $\Rz{\varphi |x|}$ and
$\Rz{\varphi x}$ on input $x = x_{n-1} \dots x_1 x_0$.
\end{lemma}

\begin{proof}
The left circuit in Figure~\ref{fig:rotations}
shows how to compute the rotation by Hamming weight.
Each input qubit controls $\Rz{\varphi}$ on the target qubit, hence the
total angle is $\varphi |x|$.  These controlled rotations are
parallelised using the parallelisation method.
\floatfigures{rotations}{Rotation by Hamming weight and value}{%
  {\def\IPEfile{pict/rotation-hamming.ipe}\begingroup
  \catcode`\%=9\catcode`\!=0\catcode`\-=11\input{\IPEfile}}
  \hskip 1cm
  {\def\IPEfile{pict/rotation-value.ipe}\begingroup
  \catcode`\%=9\catcode`\!=0\catcode`\-=11\input{\IPEfile}}
}
The right circuit shows the rotation by value.  It is similar to the
rotation by Hamming weight, only the input qubit $\ket{x_j}$ controls
$\Rz{\varphi 2^j}$, hence the total angle is $\varphi \sum_{j=0}^{n-1}
2^j x_j = \varphi x$.
\end{proof}

\begin{remark}
The construction uses rotations $\Rz{\varphi}$ for arbitrary $\varphi
\in \R$.  However, we are only allowed to use a fixed set of one-qubit
gates.  It is easy to see that every rotation can be approximated with
polynomially small error by $\Rz{\theta q} = \left( \Rz{\theta} \right)^q$,
where $\sin \theta = \frac 3 5$ and $q$ is a polynomially large
integer~\cite{adh:qcomputability}.
These $q$ rotations commute, so can be applied in parallel and the depth is
preserved.
The approximation can be kept down to polynomially small error while
increasing the size of the circuit only polynomially.
\end{remark}

\section{Constant-depth approximate circuits}

\subsection{Or gate}

It is easy to see that the rotation by Hamming weight of a string $y$ of
length $m$ with angle $\varphi = {2 \pi \over m}$ can be used to
distinguish the zero string $y = 0^m$ from strings with approximately
$\frac m2$ ones.  We, however, want to distinguish the zero string from
{\em all} nonzero strings.  It turns out that if we compute $m = \Oh{n
\log n}$ rotations by Hamming weight of the input $x$ with angles
distributed evenly around the circle, we obtain a string $y$ that is
either zero (for $x = 0^n$), or has expected Hamming weight $\frac m 2$
(for $x \ne 0^n$).  By combining these two results, we can approximate the
Or gate and, with a minor modification, also the exact[t] gate in
constant depth.

Let $w \in \N_0$ and let $\varphi$ be an angle.
Define a notation for the following one-qubit state:
\begin{equation}
\label{eq:rotate}
\Rotate \varphi w
= (H \cdot \Rz{\varphi w} \cdot H)\, \ket0
= {1 + e^{i \varphi w} \over 2} \ket0
	+ {1 - e^{i \varphi w} \over 2} \ket1.
\end{equation}
By Lemma~\ref{lem:rotation}, $\Rotate \varphi {|x|}$ can be computed in
constant depth and linear size.

\begin{theorem}
\label{th:or}
Or $\in \Rc \QNCwf^0$.  In particular, Or can be approximated with one-sided 
error $\frac1n$ in constant depth and size $\Oh{n^2 \log n}$.
\end{theorem}

\begin{proof}
Let $n$ denote the size of the input $x$.
Let $m = a \cdot n$, where $a$ will be chosen later.  For all $k \in \{ 0, 1,
\dots, m-1 \}$, compute in parallel $\ket{y_k} = \Rotate {\varphi_k} {|x|}$
for angle $\varphi_k = {2 \pi \over m} k$.  If $\ket{y_k}$ is measured in the
computational basis, the expected value of the outcome $Y_k \in \{ 0, 1 \}$ is
\[
\newcommand{\xx}[1] {e^{#1 i \varphi_k |x|}}
E[Y_k] = \left| {1 - \xx{} \over 2} \right|^2 =
	\left| \xx{-} \right| \cdot
		{\left| \xx{} + \xx{-} - 2 \right| \over 4} =
	{1 - \cos (\varphi_k |x|) \over 2}.
\]
If all these $m$ qubits $\ket{y}$ are measured, the expected Hamming weight of
all $Y$'s is
\[
E [|Y|] =
E \left[ \sum_{k=0}^{m-1} Y_k \right] = \frac m2 - \frac 12 \sum_{k=0}^{m-1}
	\cos \left( {2 \pi k \over m} |x| \right)
= \left\{ \begin{tabular}{ll}
	0 & if $|x| = 0$, \\
	$m \over 2$ & if $|x| \not= 0$. \\
	\end{tabular} \right.
\]
The qubits $\ket{y}$ are actually not measured, but their Hamming weight
$|y|$ controls another rotation on a new ancilla qubit $\ket{z}$.  So compute
$\ket{z} = \Rotate {2 \pi / m} {|y|}$.
Let $Z$ be the outcome after $\ket{z}$ is measured.  If $|y| = 0$, then
$Z = 0$ with certainty.  If $\left| |y| - \frac m2 \right| \le {m \over
\sqrt n}$, then
\[
P[Z=0] = \left| {1 + e^{i {2 \pi \over m} |y|} \over 2} \right|^2 =
{1 + \cos \left( {2 \pi \over m} |y| \right) \over 2} \le
{1 - \cos {2 \pi \over \sqrt n} \over 2} = \Oh{1 \over n}.
\]
Assume that $|x| \neq 0$.  We want to upper-bound the probability of the
bad event that $|Y|$ is not close to $\frac m2$.
Since $0 \le Y_k \le 1$, we can use
Hoeffding's Lemma~\ref{lem:hoeffding} below and obtain
\(
P \big[ \left| |Y| - \frac m2 \right| \ge {\varepsilon m} \big]
	\le {1 \over 2^{\varepsilon^2 m}}.
\)
Fix $a = \log n$ and $\varepsilon = {1 \over \sqrt n}$.  Now, 
\(
P \big[ \left | |y| - \frac m2 \right| \ge {m \over \sqrt n} \big] \le 
{1 \over 2^{m / n}} = {1 \over 2^a} = {1 \over n}.
\)
The probability that we observe the incorrect result $Z=0$ is at most
the sum of the probabilities of the two bad events, i.e. $\Oh{\frac
1n}$.  Hence
\[
P[Z=0] = \left\{ \begin{tabular}{ll}
	$1$ & if $|x| = 0$, \\
	$\Oh{1 \over n}$ & if $|x| \not= 0$. \\
	\end{tabular} \right.
\]
The circuit has constant depth and size $\Oh{m n} = \Oh{n^2 \log n}$.
It is outlined in Figure~\ref{fig:or-approx}.  The figure is slightly
simplified: unimportant qubits and uncomputation of ancillas are
omitted.
\end{proof}
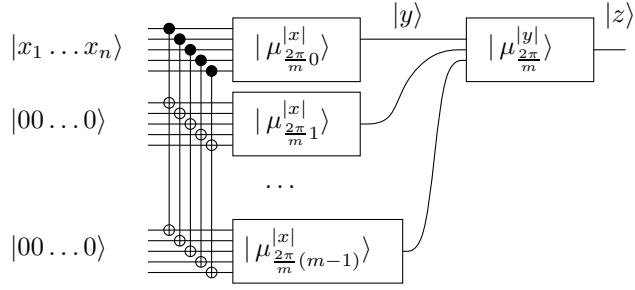
\floatfigure{or-approx}{Constant depth circuit approximating Or}

\begin{lemma}[Hoeffding
\refers{\cite{hoeffding:probineq}}]
\label{lem:hoeffding}
If $Y_1, \dots, Y_m$ are independent random variables bounded by $a_k
\le Y_k \le b_k$, then, for all $\varepsilon > 0$,
\[
P \left[ \left| S - E[S] \right| \ge \varepsilon m \right] \le
2 \exp {-2 m^2 \varepsilon^2 \over \sum_{k=1}^m (b_k - a_k)^2},
\quad \mbox{where $S = \sum_{i=k}^m Y_k$.}
\]
\end{lemma}

\begin{remark}
Since the outcome $z$ of the circuit in Figure~\ref{fig:or-approx} is a
classical bit, we can save it in an ancilla qubit by applying a controlled-not
gate and clean $\ket y$ by uncomputation.  It remains to prove that the
intermediate qubits $\ket{y}$ need not be measured, in order to be able to
uncompute them.
We show above that the output qubit is a good approximation of the
logical Or, provided $\ket{y}$ is immediately measured.  By the principle of
deferred measurement, we can use controlled quantum operations and
measure $\ket{y}$ at the end.  However, the output bit is close to a
classical bit (the distance depends on the error of the computation),
thus it is only slightly entangled with $\ket{y}$, and hence it does not
matter whether $\ket{y}$ is measured.
\end{remark}

\begin{definition}
Let $\ilog k x$ denote the $k$-times \emph{iterated logarithm} $\log \log
\dots \log x$.  The \emph{log-star function}, $\logstar x$, is the maximum
number of iterations $k$ such that $\ilog k x$ exists and is real.%
\footnote{The log-star of the estimated number of atoms in the universe is 5.
Consequently, for the computational problems we consider in this paper, the
log-star is in practice at most 5.}
\end{definition}

\begin{remark}
If we require error $1 \over n^c$, we create $c$ copies and compute
the exact Or of them by a binary tree of Or gates.  The tree has depth $\log
c = \Oh{1}$.
In Section~\ref{sec:small-size-or}, we show how to approximate Or
in constant depth and size $\OO{n \ilog k n}$ for any constant $k$.  In
Section~\ref{sec:linear-size}, we show how to compute Or {\em exactly} in log-star
depth and linear size.
\end{remark}

\subsection{Exact[t] and threshold[t] gates}

\begin{theorem}
\label{th:exactt-1}
exact[t] $\in \Rc \QNCwf^0$.
\end{theorem}

\begin{proof}
We slightly modify the circuit for Or.  As outlined in
Figure~\ref{fig:rotation-exactt},
by adding the rotation $\Rz{- \varphi t}$ to the rotation by Hamming weight
in the first layer, we obtain $\Rotate \varphi {|x|-t}$
instead of $\Rotate \varphi {|x|}$.  The second layer stays the same.
If the output qubit $\ket{z}$ is measured, then
\[
P[Z=0] = \left\{ \begin{tabular}{ll}
	$1$ & if $|x| = t$, \\
	$\Oh{1 \over n}$ & if $|x| \not= t$. \\
	\end{tabular} \right.
\]
We obtain an approximation of the exact[t] gate with one-sided
polynomially small error.
\end{proof}
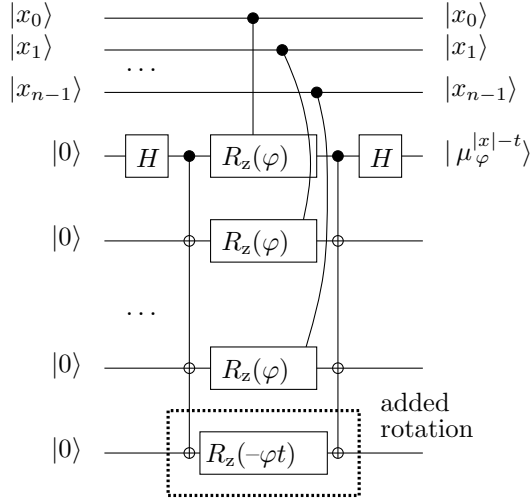
\floatfigure{rotation-exactt}{Rotation by Hamming weight with added
rotation}

\begin{remark}
Other gates are computed from the exact[t] gate by standard methods.  For example,
threshold[t] can be computed as the parity of exact[$t$], exact[$t+1$],
\dots, exact[$n$].  The depth stays constant and the size is just
$n$-times bigger, i.e.\ $\Oh{n^3 \log n}$, hence threshold[t] $\in \Bc
\QNCwf^0$.
In Section~\ref{sec:small-size-counting}, we show how to
approximate exact[t], threshold[t], and counting in constant depth and size
$\Oh{n \log n}$.
\end{remark}

\subsection{Arithmetic operations}

Using threshold gates, one can do arithmetic operations in constant depth.  The
following circuits take as part of the input an ancilla register in state
$\ket 0$ and output the result of the computation in that register.

\begin{theorem}
\label{th:arithmetic}
The following functions are in $\Bc \QNCwf^0$:
addition and multiplication of $n$ integers,
division of integers with remainder,
and sorting of $n$ integers.
\end{theorem}

\begin{proof}
By~\cite{sbkh:division-circ}, these functions are computed by constant-depth,%
\footnote{The depths are really small, from 2 to 5.}
polynomial-size threshold circuits.  A threshold circuit is built of
weighted threshold gates.
It is simple to prove that the weighted threshold gate (with
polynomially large integer weights) also is in $\Bc \QNCwf^0$.  One only
needs to rotate the phase of the quantum state in
Lemma~\ref{lem:rotation} by integer multiples of the basic angle.
\end{proof}

In the following section, we require a reversible version of modular
addition.

\begin{definition}
Let $q$ be an $n$-bit integer and $x_1, \dots, x_m \in \Z_q$.
The \emph{reversible addition} gate
maps $\add^m: \ket q \ket{x_1} \dots \ket{x_m} \to \ket q
\ket{x_1} \dots \ket{x_{m-1}} \ket{y}$,
where $y = \left( \sum_{i=1}^m x_i \right) \mod q$.
\end{definition}

\begin{lemma}
\label{lem:add}
$\add^m \in \Bc \QNCwf^0$.
\end{lemma}

\begin{proof}
By Theorem~\ref{th:arithmetic}, $y = (\sum_{i=1}^m x_i) \mod q$ can be
approximated in constant depth and polynomial size.  The result is, however,
stored into ancilla qubits.  Hence we have to erase $x_m$, which we may
achieve by first negating the contents in $y$ by $\ket{y} \rightarrow
\ket{-y}$, computing the sum $w = y + \sum_{i=1}^{m-1} x_i$ in a fresh
ancilla, do a bitwise control-not of $w$ into $x_m$, uncompute $w$, and
finally re-negate $y$.  We then swap the ancillas $\ket{y}$ with the erased
qubits in $\ket{x_m}$.
\end{proof}

\subsection{Quantum Fourier transform}

The QFT is a very powerful tool used in several quantum algorithms, e.g.\ 
factoring of integers and computing the discrete logarithm~\cite{shor:factoring}.

\begin{definition}
\label{def:qft}
The \empdef{quantum Fourier transform} with respect to modulus $q$ performs
the Fourier transform on the quantum amplitudes of the state, i.e.\ it maps
\begin{equation}
F_q: \ket{x} \to \Phase x  = {1 \over \sqrt{q}}
	\sum_{y=0}^{q-1} \omega^{x y} \ket{y}
\mbox{, where $\omega = e^{2 \pi i /q}$,}
\end{equation}
for $x \in \{ 0, 1, \dots, q-1 \}$ and it behaves arbitrarily on the other
states.
\end{definition}

\subsubsection{QFT with a power-of-2 modulus}

Let $q = 2^n$.
Coppersmith has shown in~\cite{coppersmith:qft} how to compute the QFT in quadratic
depth, quadratic size, and without ancillas.  The depth has further been
improved to linear [folklore].  Cleve and Watrous have shown in~\cite{cw:qft} that
the QFT can be approximated with error $\varepsilon$ in depth $\Oh{\log n +
\log \log {1 \over \varepsilon}}$ and size $\Oh{n \log {n \over \varepsilon}}$.
They also show that if only gates acting on a constant number of qubits are
allowed (in particular, the fan-out gate is not allowed), logarithmic depth is
necessary.  We show that the approximate circuit for the QFT
from~\cite{cw:qft} can be compressed to constant depth, if we allow the
fan-out gate.

\begin{theorem}
\label{th:qft}
QFT $\in \Bc \QNCwf^0$.
\end{theorem}

\begin{proof}
The operator $F_{2^n}: \ket{x} \to \Phase x$ can be computed by composing:
\[
\begin{tabular}{l l l l}
1. & Fourier state construction (QFS):
	& $\ket{x} \ket0 \dots \ket0$
	& $\to \ket{x} \Phase x \ket0 \dots \ket0$ \\
2. & Copying Fourier state (COPY):
	& $\ket{x} \Phase x \ket0 \dots \ket0$
	& $\to \ket{x} \Phase x \dots \Phase x$ \\
3. & Uncomputing phase estimation (QFP):
	& $\Phase x \dots \Phase x \ket{x}$
	& $\to \Phase x \dots \Phase x \ket0$ \\
4. & Uncomputing COPY:
	& $\Phase x \dots \Phase x \ket 0$
	& $\to \Phase x \ket0 \dots \ket0$ \\
\end{tabular}
\]
The following lemmas show that each of these individual operators is in $\Bc
\QNCwf^0$.
\end{proof}

\begin{lemma}
\label{lem:qfs}
QFS $\in \QNCwf^0$.
\end{lemma}

\begin{proof}
QFS maps $\ket{x} \ket0 \to \ket{x} \Phase x$.
Define $\phase r = {\ket0 + e^{2\pi i r} \ket1 \over \sqrt 2}$.
It is simple to prove that $\Phase x = \QFTbit 1 \QFTbit 2 \dots
\QFTbit n$.
\begin{eqnarray*}
\Phase x &=& {1 \over \sqrt{2^n}} \sum_{y=0}^{2^n-1} \omega^{x y} \ket{y}
= {1 \over \sqrt{2^n}} \sum_{y=0}^{2^n-1}
	\bigotimes_{k=1}^n \omega^{x 2^{n-k} y_{n-k}} \ket{y_{n-k}}
\\
&=& {1 \over \sqrt{2^n}} \bigotimes_{k=1}^n \sum_{b=0}^1
	(\omega^{2^{n-k} x})^b \ket{b}
= \bigotimes_{k=1}^n {\ket{0} + e^{2 \pi i x / 2^k} \ket{1} \over \sqrt2}
= \bigotimes_{k=1}^n \phase{x / 2^k}.
\end{eqnarray*}
The $n$ qubits $\QFTbit k$ can be computed from $x$ in
parallel as follows:  $\QFTbit k = \Rz{{2 \pi \over 2^k} x} {\ket0 +
\ket1 \over \sqrt2}$ is computed by the rotation by value
(Lemma~\ref{lem:rotation}) in constant depth and linear size.
\end{proof}

\goodbreak
\begin{lemma}
\label{lem:copy}
COPY $\in \Bc \QNCwf^0$.
\end{lemma}

\begin{proof}
COPY maps $\Phase x \ket0 \dots \ket0 \to \Phase x \dots \Phase x$.
Take the reversible addition gate modulo $2^n$: $(\add^2_{2^n}\!) \ket{y} \ket{x} = \ket{y}
\ket{(x + y) \mod 2^n}$.  It is simple to prove that $\add^{-1}
\Phase y \Phase x = \Phase {x+y} \Phase x$.
\begin{eqnarray*}
\Phase y \Phase x
&=& \displaystyle {1 \over 2^n} \sum_{l,k=0}^{2^n-1} \omega^{l y+ k x} \ket{l} \ket{k}
\to_{\add^{-1}} {1 \over 2^n} \sum_{l,k} \omega^{l y + k x} \ket{l} \ket{k-l}
\\
&=& \displaystyle {1 \over 2^n} \sum_{l,m} \omega^{l y + (m+l)x} \ket{l} \ket{m}
= {1 \over 2^n} \sum_{l,m} \omega^{l(x+y) + m x} \ket{l} \ket{m}
= \Phase {x+y} \Phase x.
\\
\end{eqnarray*}
Hence $\add^{-1}
\Phase 0 \Phase x = \Phase x \Phase x$.  The state $\Phase 0 =
H^{\otimes n} \ket{0^n}$ is easy to prepare in constant depth.
Furthermore, $(\add^m_{2^n}\!)^{-1} \Phase 0 \dots \Phase 0 \Phase x = \Phase x
\dots \Phase x \Phase x$, because the addition of $m-1$ numbers into one
register is equivalent to $m-1$ consecutive additions of one number.
Each such a reversible addition copies $\Phase x$ into 1 register.  Note
that the $\add^m_{2^n}$ gate performs all these additions in parallel.  By
Lemma~\ref{lem:add}, the reversible addition gate is in $\Bc \QNCwf^0$.
\end{proof}

\begin{lemma}
QFP $\in \Bc \QNCwf^0$.
\end{lemma}

\begin{proof}
QFP maps $\Phase x \dots \Phase x \ket0 \to \Phase x \dots \Phase x \ket{x}$.
By Cleve and Watrous \cite[Section 3.3]{cw:qft}, we can compute $x$ with
probability at least $1 - \varepsilon$ from $\Oh{\log {n \over \varepsilon}}$
copies of $\Phase x$ in depth $\Oh{\log n + \log \log {1 \over \varepsilon}}$
and size $\Oh{n \log {n \over \varepsilon}}$.  Use $\varepsilon = {1 \over
\poly(n)}$.  It is simple to convert their circuit into constant depth,
provided we have fan-out.  The details are sketched below.

The input consists of $m = \Oh{\log {n \over \varepsilon}}$ copies of $\Phase x =
\QFTbit 1 \QFTbit 2 \dots \QFTbit n$.  Measure each $\QFTbit k$ $m \over
2$ times in the basis $\{ \phase{0.01}, \phase{0.11} \}$ and $m \over 2$
times in the Hadamard basis $\{ \phase{0.00}, \phase{0.10} \}$.
The state $\QFTbit k = \frac1{\sqrt2} (\ket0 + e^{2\pi i (0.x_{k-1} \dots x_1 x_0)})$
lies on the middle circle of the Bloch sphere; it
is shown in Figure~\ref{fig:qfp-measure2}.  If
$\QFTbit k$ is in the white region, then the measurement in the first
basis tells whether $x_{k-1} = 0$ or $1$ with probability at least $3
\over 4$.  If $\QFTbit k$ is in the shaded region, then the measurement
in the Hadamard basis tells whether $x_{k-1} = x_{k-2}$ or $\neg
x_{k-2}$ (denoted by P, N) with probability at least $3 \over 4$.  
\floatfigure{qfp-measure2}{Measurement of $\QFTbit k$ in a random basis}

For each $k$, perform the majority vote and obtain the correct answer
$z_k \in \{ 0, 1, \rm P, N \}$ with error probability at most ${1
\over 2^m} = {\varepsilon \over n}$.  The probability of having any
error is at most $n$ times bigger, i.e.\ at most $\varepsilon$.  Compute
$x_{n-1} \dots x_1 x_0$ from $z_{n-1} \dots z_1 z_0$ in constant depth.
The bit $x_k$ is computed as follows:
\begin{steps}
\item If $z_k z_{k-1} \dots z_{l+1} \in \{ \rm P, N \}$ and $z_l \in \{
0, 1 \}$, compute the parity of the number of N's and add it to $z_l$
(assuming $z_{-1}=0$), otherwise return 0.
\item Check and compute all prefixes $l$ in parallel and take the logical Or of the
results.
\end{steps}
\noindent
All the gates used (fan-out, parity, And, Or, majority) are in $\Bc
\QNCwf^0$.
\end{proof}

\subsubsection{QFT with an arbitrary modulus}

Let $q \ne 2^n$.  Cleve and Watrous have shown in~\cite{cw:qft} that the QFT
can be approximated with error $\varepsilon$ in depth $\Oh{(\log \log
q)(\log \log {1 \over \varepsilon})}$ and size $\poly(\log q + \log {1
\over \varepsilon})$.  We show that their circuit can also be compressed
into constant depth, if we use the fan-out gate.  The relation between quantum
Fourier transforms
with different moduli was described in~\cite{hh:qft}.

\begin{remark}
We actually implement a slightly more general operation, when $q$ is not a
fixed constant, but an $n$-bit {\em input} number.  This generalised QFT maps
$\ket{q} \ket{x} \to \ket{q} \Phase x$.  The register $\ket{q}$ is implicitly
included in all operations.  We will henceforth omit it and the generalised
operations are denoted simply by QFT$_q$, QFS$_q$, COPY$^m_q$, and QFP$_q$.
\end{remark}

\begin{theorem}
\label{th:qftq}
QFT$_q$ $\in \Bc \QNCwf^0$.
\end{theorem}

\newcommand{\dummy}[1]{\ket{\mathrm{dummy}_{#1}}}
\begin{proof}
Let $\dummy{q,x}$ denote an unspecified quantum state depending on two
parameters $q, x$.
The operator $F'_q: \ket{x} \to \Phase x \dummy{q,0}$ can be computed by composing:
\[
\begin{tabular}{l l l l}
1. & QFS$_q$:
	& $\ket{x} $
	& $\to \ket{x} \Phase x \dummy {q,x}$ \\
2. & COPY$^{m+1}_q$:
	&& $\to \ket x \Phase x \dummy {q,x}
	{\left( \Phase x \dummy {q,0} \right)}^{\otimes m}$ \\
3. & Uncomputing QFS$_q$:
	&& $\to \ket x
	{\left( \Phase x \dummy {q,0} \right)}^{\otimes m}$ \\
4. & Uncomputing QFP$_q$:
	&& $\to
	{\left( \Phase x \dummy {q,0} \right)}^{\otimes m}$ \\
5. & Uncomputing COPY$^m_q$:
	&& $\to \Phase x \dummy {q,0}$, \\
\end{tabular}
\]
where empty registers are omitted for clarity.
The state $\dummy{q,0}$ is not entangled with $\ket{x}$ and hence it can be
traced out.  We obtain the quantum Fourier transform $F_q$.
The following lemmas show that each of these individual operators is in $\Bc
\QNCwf^0$.
\end{proof}

\begin{lemma}
\label{lem:qfsq}
QFS$_q \in \Bc \QNCwf^0$.
\end{lemma}

\begin{proof}
QFS$_q$ maps $\ket{x} \ket0 \to \ket{x} \Phase x \dummy{q,x}$ for some
``garbage'' state $\dummy{q,x}$.
We will show that QFS$_q$ is well approximated by a QFS with a
power-of-2 modulus of the magnitude $q^3$.
Let $n = \lceil \log q \rceil$.  Take $N=3n$ and extend $x$ by leading zeroes
into $N$ bits.  Using Lemma~\ref{lem:qfs}, perform QFS$_{2^N}$
and obtain the state
\(
\ket{x} {1 \over \sqrt{2^N}} \sum_{y=0}^{2^N-1}
	e^{{2 \pi i \over 2^N} x y} \ket{y}.
\)

Set $u = \lfloor {2^N / q} \rfloor$ and apply integer division by $u$ to the
second register, i.e.\ map $\ket{y} \to \ket{y_1} \ket{y_2}$, where
$y_1 = \lfloor {y / u} \rfloor \in \{ 0, 1, \dots, q \}$
and $y_2 = y \mod u$.  This can be done reversibly in
constant depth by a few applications of Theorem~\ref{th:arithmetic}
using the method from Lemma~\ref{lem:add}.  The quantum state can be
written as
\[
{1 \over \sqrt{2^N}} \sum_{y=0}^{2^N-1} e^{{2 \pi i \over 2^N} x y}
	\ket{y_1} \ket{y_2} =
{1 \over \sqrt{2^N}} \sum_{y_1=0}^{q-1} \sum_{y_2=0}^{u-1}
	e^{{2 \pi i \over 2^N} x (y_1 u + y_2)} \ket{y_1} \ket{y_2}
	+ \ket{w},
\]
where $\ket{w} = {1 \over \sqrt{2^N}} \sum_{z=0}^{v-1} e^{{2 \pi i \over
2^N} x (q u + z)} \ket{q} \ket{z}$ and $v = 2^N \mod u =
2^N - q u = 2^N \mod q < q$.  The sum has been
rearranged using $y = y_1 u + y_2$.  Now, $\norm{\ket{w}} = \sqrt{v \over
2^N} = \Oh{2^{-n}}$ is exponentially small and so it can be neglected.
Decompose the quantum state into the tensor product
\[
{1 \over \sqrt{q}} \sum_{y_1=0}^{q-1}
	e^{{2 \pi i \over q} ({q \over 2^N}u) x y_1} \ket{y_1} 
	\otimes \sqrt{q \over 2^N} \sum_{y_2=0}^{u-1} 
	e^{{2 \pi i \over 2^N} x y_2} \ket{y_2}.
\]
Now, $u$ is exponentially close to $2^N \over q$, because ${q \over 2^N} u =
{2^N - v \over 2^N} = 1 - \Oh{2^{-2n}}$.  Since ${x y_1 \over q} = \Oh{2^n}$,
the replacement of $q u \over 2^N$ by 1 in the exponent causes only
exponentially small error $\Oh{2^{-n}}$.  Hence the quantum state is exponentially close to
\[
{1 \over \sqrt{q}} \sum_{y=0}^{q-1} e^{{2 \pi i \over q} x y} \ket{y}
	\otimes {1 \over \sqrt{u}} \sum_{z=0}^{u-1}
	e^{{2 \pi i \over 2^N} x z} \ket{z}
= \Phase x \dummy{q,x}.
\]
The ``garbage'' state $\dummy{q,x}$ arises as a byproduct of the higher
precision $3 n$-bit arithmetic.  We clean it up later by uncomputing QFS$_q$
after copying $\Phase x$; see the proof of Theorem~\ref{th:qftq}.  It
actually gets replaced by $\dummy{q,0} = {1 \over \sqrt u}
\sum_{z=0}^{u-1} \ket z$, which does not depend on $x$ and it thus
causes no harm.  We have approximated QFS$_q$ in constant depth.
\end{proof}

\begin{lemma}
COPY$^m_q \in \Bc \QNCwf^0$.
\end{lemma}

\begin{proof}
COPY$^m_q$ maps $\Phase x \ket0 \dots \ket0 \to
\Phase x (\Phase x \dummy{q,0})^{\otimes (m-1)}$.
The proof is similar to the proof of Lemma~\ref{lem:copy}.  First, prepare $m-1$
states $\Phase 0 \dummy{q,0}$ by applying QFS$_q$ to $\ket0 \ket0$
(Lemma~\ref{lem:qfsq}).  Second, use the inverse of the reversible addition modulo $q$ to
map $(\add^m_q)^{-1}: \Phase 0 \dots \Phase 0 \Phase x \to \Phase x \dots
\Phase x \Phase x$ (Lemma~\ref{lem:add}).
\end{proof}

\begin{lemma}
QFP$_q \in \Bc \QNCwf^0$.
\end{lemma}

\begin{proof}
QFP$_q$ maps $\Phase x \dots \Phase x \ket0 \to \Phase x \dots \Phase x \ket{x}$.
We use an idea similar to the proof of Lemma~\ref{lem:qfsq}.  Let $n =
\lceil \log q \rceil$ and $N=3n$.  Extend $\Phase x$ by leading zeroes to $N$ bits and
apply $F^\dagger_{2^N}$ to them (Theorem~\ref{th:qft}).  We obtain many copies
of the state
\[
F^\dagger_{2^N} (\ket0 \Phase x) = {1 \over \sqrt{2^N q}}
	\sum_{z=0}^{2^N-1} \left( \sum_{y=0}^{q-1}
	e^{{-2\pi i \over 2^N} z y + {2 \pi i \over q} x y} \right)
	\ket{z}.
\]
The exponent can be rewritten to $2 \pi i ({x \over q} - {z \over 2^N})
\cdot y$.  Intuitively, if $|z - 2^N {x \over q}| \le {2^N \over 8q}$,
then $|\frac x q - \frac z {2^N}| \le \frac 1 {8q}$, the absolute value
of the angle in the exponent is at most $\frac \pi 4$ for every $y \in
\{ 0, 1, \dots, q-1 \}$, and the amplitudes sum up constructively.  If
$z$ is not close to $2^N {x \over q}$, then the amplitudes interfere
destructively.  The quantum state has most of its amplitude on the good
$z$'s.  So we compute reversibly by division with remainder an estimate
$x' = \lfloor {z q \over 2^N} + \frac12 \rfloor$.  A detailed analyzis
shows that $P[x'=x] \ge \frac12 + \delta$ for some constant $\delta$
\cite{cw:qft,hh:qft}.  Here we do not present the details, because our
goal is the compression of the circuit from~\cite{cw:qft} into constant
depth.

We transform all $m = \Oh{\log {n \over \varepsilon}}$ input quantum
states $\Phase x$ into $m$ independent estimates $\ket{x'}$.  We then
estimate all bits of $x$ one-by-one from these $m$ estimates by majority
gates.  Each bit of $x$ is wrong with probability at most $2^{-m} = 2^{-\log
{n \over \varepsilon}} = {\varepsilon \over n}$.  The probability of
having an error among the $n$ bits of $x$ is thus at most $\varepsilon$.
Finally, save the estimation of $x$ in the target register and uncompute the
divisions and the quantum Fourier transforms.  With probability at least
$1-\varepsilon$, the mapping QFP$_q$ has been performed.  Use
$\varepsilon = {1 \over \poly{n}}$.
\end{proof}

\subsection{Quantum phase estimation}

The method of computing QFT$_{2^n}$ can be also used for phase estimation.

\begin{theorem}
Given a gate $S_x: \ket{y} \ket{\phi} \to \ket{y} \Rz{{2 \pi x \over
2^n} y} \ket{\phi}$ for basis states $\ket{y}$, where $x \in \Z_{2^n}$ is
unknown, we can determine $x$ with probability at least $1 - \varepsilon$ in
constant depth, size $\Oh{n \log {n \over \varepsilon}}$, and using the
$S_x$ gate $\Oh{n \log {n \over \varepsilon}}$ times.
\end{theorem}

\begin{proof}
Obtain an estimate of $x$ by applying the QFP to $\Oh{\log {n \over \varepsilon}}$
copies of the quantum state $\Phase x = \QFTbit 1 \QFTbit 2 \dots \QFTbit n$.
Each $\QFTbit k$ can
be computed by one application of $S_x$ to  $\ket{2^{n-k}} {\ket{0} + \ket1
\over \sqrt2}$, because $\QFTbit k = \Rz{2 \pi x \over 2^k} {\ket0 + \ket1
\over \sqrt2} = \Rz{{2 \pi x \over 2^n} 2^{n-k}} {\ket0 + \ket1 \over
\sqrt2}$.
\end{proof}

\section{Exact circuits of small depth}
\label{sec:log-star}

In the previous section, we have shown how to {\em approximate} the exact[t]
gate in constant depth.  In this section, we show how to {\em compute it
exactly} in log-star depth.  The circuits in this section use arbitrary
one-qubit gates instead of a fixed basis, otherwise they would not be exact.

\begin{lemma}
\label{lem:reduction}
The function Or on $n$ qubits can be reduced {\em exactly} to Or on $m = \lceil \log
(n+1) \rceil$ qubits in constant depth and size $\Oh{n \log n}$.
\end{lemma}

\begin{proof}
We use a technique similar to the proof of Theorem~\ref{th:or}.
Recall the quantum state $\Rotate \varphi w$ defined by
equation~(\ref{eq:rotate}) on page~\pageref{eq:rotate}.
For $k \in \{ 1, 2, \dots, m \}$, compute in parallel $\ket{y_k} = \Rotate
{\varphi_k} {|x|}$ for angle $\varphi_k = {2 \pi \over 2^k}$.
Let $\ket y = \ket{y_1 y_2 \dots y_m}$.
\begin{itemize}
\item If $|x| = 0$, then $\braket y {0^m} = 1$, because $\ket{y_k} = \ket0$ for each $k$.

\item If $|x| \neq 0$, then $\braket y {0^m} = 0$, because at least one
qubit $y_k$ is one with certainty.  Take the unique decomposition of
$|x|$ into a product of a power of 2 and an odd number: $|x| = 2^a (2b + 1)$
for $a, b \in \N_0$.  Then
\[
\braket 1 {y_{a+1}} =
{1 - e^{i \varphi_{a+1} |x|} \over 2} =
{1 - e^{i {2 \pi \over 2^{a+1}} 2^a (2b + 1)} \over 2} =
{1 - e^{i \pi (2b +1)} \over 2} =
{1 - e^{i \pi} \over 2} = 1.
\]
\end{itemize}
It follows that $x$ is non-zero if and only if $y$ is.  Hence the
original problem is exactly reduced to a problem of logarithmic size.
\end{proof}

\begin{theorem}
\label{th:exactt-ac1}
exact[t] $\in \QACwf^0$.
\end{theorem}

\begin{proof}
Using the methods from Theorem~\ref{th:exactt-1} and
Lemma~\ref{lem:reduction}, exact[t] can also be reduced to Or of
logarithmic size.  The reduction has constant depth and size $\Oh{n \log
n}$.  Hence exact[t] is $\QNCwf^0$-reducible to Or, or simply exact[t]
$\in \QACwf^0$, because $\QACwf^0$ includes both $\QNCwf^0$ and the Or
gate.
\end{proof}

\begin{theorem}
\label{th:exactt-logstar}
exact[t] $\in \QNCwf(\logstar n)$, i.e.\ exact[t] can be computed {\em
exactly} in log-star depth and size $\Oh{n \log n}$.
\end{theorem}

\begin{proof}
Apply the reduction used in Lemma~\ref{lem:reduction} in total $(\logstar
n)$-times, until the input size is at most $2$.  Compute and save the outcome,
and clean ancillas by uncomputation.  The circuit size is $\Oh{n \log n}$.
\end{proof}

\section{Circuits of small size}

In this section, we decrease the size of some circuits.  We
allow the use of arbitrary one-qubit gates instead of a fixed basis.

\subsection{Constant depth approximation of Or}
\label{sec:small-size-or}

In this section, we apply the reduction from Lemma~\ref{lem:reduction}
repeatedly to shrink the circuit for Or.  We first reduce the size of the circuit
to $\Oh{n \log n}$.  We then develop a recurrent method that reduces the size
even further.  Let us define a useful notation.

\begin{definition}
Let $x = x_1 x_2 \dots x_n$.  By {\em Or-reduction} $n \to m$ with error
$\varepsilon$ we mean a quantum circuit mapping $\ket x \ket {0^m} \to \ket x
\ket \varphi$ such that, if $|x|=0$, then $\ket \varphi = \ket {0^m}$ and, if
$|x| \ne 0$, then $\braket {0^m} \varphi \le \varepsilon$.
\end{definition}

The Or-reduction preserves the logical Or of qubits, i.e.\ $|x|=0$ iff $|\varphi|=0$ with
high probability.  Theorem~\ref{th:or} provides an Or-reduction $n \to 1$ with
error $\frac1n$, constant depth, and size $n^2 \log n$.
Lemma~\ref{lem:reduction} provides an Or-reduction $n \to \log n$ with error
$0$, constant depth, and size $n \log n$.

\begin{lemma}
\label{lem:nlogn-approx}
There is an Or-reduction $n \to 1$ with error $\frac1n$, constant depth, and
size $n \log n$.
\end{lemma}

\begin{proof}
Divide the input into $\sqrt n \over \log n$ blocks of size $\sqrt n \log n$.
First, reduce each block by Lemma~\ref{lem:reduction} to $\frac12 \log n +
\log \log n = \Oh{\log n}$ qubits in constant depth and size $\sqrt n \log^2
n$.  In total, we obtain $\sqrt n$ new qubits in size $n \log n$.  Second,
compute the logical Or by Theorem~\ref{th:or} in constant depth, size ${\sqrt n}^2 \log
\sqrt n = \Oh{n \log n}$, and error $1 \over \sqrt n$.  To amplify the error
to $1 \over n$, repeat the computation twice and return 1 if any of them
returns 1 (the error is one-sided).  The circuit size is doubled.
\end{proof}

The circuit size can be reduced to $\OO{n \log^{(d)} n}$ for
any constant number $d$ of iterations of the logarithm.  The trick is to
divide input qubits into small blocks and perform the reduction step on
each of them.  The number of variables is reduced by a small factor and
we can thus afford to apply a circuit of a slightly bigger size.  It we
repeat this reduction step $d$ times, we obtain the desired circuit.

\begin{theorem}
\label{th:ilogn-approx}
There exist constants $c_1, c_2$ such that for every $d \in \N$, there is an
Or-reduction $n \to 1$ with error $\frac1n$, depth $c_1 d$, and size $c_2 d n
\ilog d n$.
\end{theorem}

\begin{proof}
By induction on $d$: we have already verified the case $d=1$ in
Lemma~\ref{lem:nlogn-approx}.
For the induction step: Divide $n$ input qubits into $n /
\ilog{d-1} n$ blocks of $\ilog{d-1} n$ qubits.  Using
Lemma~\ref{lem:reduction}, reduce each block to $\ilog d n$ qubits in constant
depth and size $c_2 \ilog{d-1} n \cdot \ilog d n$.  Total size is $c_2 n \ilog
d n$.  We obtain ${n \over \ilog{d-1} n} \ilog d n = o(n)$ new qubits.  Using
the induction hypothesis, compute their logical Or in depth $c_1 (d-1)$ and size $c_2
(d-1) \left({n \over \ilog{d-1} n} \ilog d n \right) \cdot \ilog{d-1} o(n) \le c_2
(d-1) n \ilog d n$.  Together, it takes depth $c_1 d$ and size $c_2 d n \ilog
d n$.

The only approximate step is the final application of
Lemma~\ref{lem:nlogn-approx} for $d=1$.  It is applied on ${n \over \log n}
\ilog d n$ variables, hence the error is $\Oh{\log n / n}$.  It can be
amplified to $1 \over n$ by running the computation twice.
\end{proof}

\subsection{Log-star depth computation of Or}
\label{sec:linear-size}

Our best constant-depth circuit for Or is described by
Theorem~\ref{th:ilogn-approx}.  It is approximate and it has slightly
super-linear size.  In this section, we show that we can achieve an {\em
exact} circuit of {\em linear} size if we relax the restriction of
constant depth.  We consider $d$ in Theorem~\ref{th:ilogn-approx} a
slowly growing function of $n$ instead of a constant.  Now we can use an
Or-reduction better than Lemma~\ref{lem:nlogn-approx}.
Theorem~\ref{th:exactt-logstar} provides an Or-reduction $n \to 1$ with
error $0$, log-star depth, and size $n \log n$.

\begin{lemma}
\label{lem:ilogn-exact}
There exist constants $c_1, c_2$ such that for every $d \in \N$, there is an
Or-reduction $n \to 1$ with error $0$, depth $c_1 d + \logstar n$, and size
$c_2 d n \ilog d n$.
\end{lemma}

\begin{proof}
The same as of Theorem~\ref{th:ilogn-approx}, but use the Or-reduction from
Theorem~\ref{th:exactt-logstar} instead of Lemma~\ref{lem:nlogn-approx} in the
last layer (for $d=1$).  The size stays roughly the same, the circuit becomes
exact, and the depth is increased by an additional term of $\log^* n$.
\end{proof}

\begin{theorem}
There is an Or-reduction $n \to 1$ with error $0$, log-star depth, and linear
size.
\end{theorem}

\begin{proof}
Divide the input into $n \over \log^* n$ blocks of size $\log^* n$.  Compute
the logical Or of each block by a balanced binary tree of depth $\log(\log^* n) <
\log^* n$ and in {\em linear} size.  Using Lemma~\ref{lem:ilogn-exact} with
$d = \logstar n$, compute the logical Or of $n \over \log^* n$ new qubits in log-star
depth and size $\Oh{\logstar n \cdot {n \over \logstar n} \cdot \log^{(\logstar
n)} n} = \Oh n$.
\end{proof}

\subsection{Approximation of counting and threshold[t]}
\label{sec:small-size-counting}

In this section, we use the QFT for the parallelisation of increments.  This
allows us to approximate the Hamming weight of the input in smaller
size $\Oh{n \log n}$.

\newcommand{\Incr}{\mathrm{Incr}}
\begin{definition}
The \empdef{increment gate} maps
\( \Incr_n: \ket{x} \to \ket{(x + 1) \mod 2^n}. \)
\end{definition}

\begin{lemma}
\label{lem:incr}
The increment gate is diagonal in the Fourier basis and its diagonal
form is in $\QNC^0$.
\end{lemma}

\begin{proof}
Let $\omega = e^{2 \pi i / 2^n}$ and let $\ket x$ be any computational basis
state.  It is simple to prove the following two equations:
\begin{steps}
\item $\Incr_n = F_{2^n}^\dagger D_n F_{2^n}$ for diagonal
$D_n = \sum_{y=0}^{2^n-1} \omega^y \ket{y} \bra{y}$.
\[
F^\dagger D F \ket{x} =
F^\dagger D {\sum_{y=0}^{2^n-1} \omega^{x y} \ket{y} \over \sqrt{2^n}} =
F^\dagger {\sum_{y=0}^{2^n-1} \omega^{(x+1)y} \ket{y} \over \sqrt{2^n}} =
\ket{(x+1) \mod 2^n}.
\]
\item $D = \Rz{\pi} \otimes \Rz{\pi / 2} \otimes
\dots \otimes \Rz{\pi / 2^{n-1}}$.
\begin{eqnarray*}
D \ket{x} &=& \omega^x \ket{x} =
\bigotimes_{k=1}^n \omega^{2^{n-k} x_{n-k}} \ket{x_{n-k}} =
\bigotimes_{k=1}^n (e^{2 \pi i / 2^k})^{x_{n-k}} \ket{x_{n-k}} =
\\
&=& \bigotimes_{k=1}^n \Rz{2 \pi / 2^k} \ket{x_{n-k}}
= \left( \Rz{\pi} \otimes \dots \otimes \Rz{\pi / 2^{n-1}} \right) \ket{x}.
\\
\end{eqnarray*}
\end{steps}
\noindent
We conclude that $\Incr = F^\dagger D F$, and that $D$ is a tensor product of
one-qubit operators.
\end{proof}

\goodbreak

\begin{remark}
The addition of a fixed integer $b$ is as hard as the
increment.  By Lemma~\ref{lem:incr}, $\Incr^b = F^\dagger D^b F$ and
$\left( \Rz{\varphi} \right)^b = \Rz{\varphi b}$, hence the diagonal
version of the addition of $b$ is also in $\QNC^0$.
\end{remark}

\begin{theorem}
\label{th:counting}
Counting can be approximated in constant depth and size $\Oh{n \log n}$.
\end{theorem}

\begin{proof}
Compute the Hamming weight of the input.  Each input qubit controls one
increment on an $m$-qubit counter initialised to $0$, where $m = \lceil
\log (n+1) \rceil$.  The increments $\Incr_m$ are parallelised
(Theorem~\ref{th:parallel} and Lemma~\ref{lem:incr}), so we apply the
quantum Fourier transform $F_{2^m}$ twice (Theorem~\ref{th:qft}) and the $n$
constant-depth controlled $D_m$ gates in parallel.  The size is
$\Oh{\poly(m) + nm} = \Oh{n \log n}$.
\end{proof}

\begin{remark}
threshold[t] is equal to the most significant qubit of the counter if we
align it to a power of 2 by adding a fixed integer $2^m-t$.  exact[t] can be
computed by comparing the counter with $t$.
\end{remark}

\section{Concluding remarks}

\subsection{Comparison with randomised circuits}

Let us compare our results for quantum circuits with similar results for
classical randomised circuits.  We consider randomised circuits with
bounded fan-in of Or and And gates, and unbounded fan-out and parity
(similar to the quantum model).  Classical lower bounds are folklore and
we attach the proofs for the convenience of the reader in
Appendix~\ref{app:lower-bounds}.

\[{
\setlength{\tabcolsep}{0.5em}
\begin{tabular}{|l|l|l|}
\hline
\bf Gate & \bf Randomised & \bf Quantum \cr
\hline
Or and threshold[t] exactly		& $\Th{\log n}$		& $\Oh{\logstar n}$ \cr
mod[q] exactly				& $\Th{\log n}$		& $\Th{1}$ \cr
Or with error $\frac 1n$		& $\Th{\log \log n}$	& $\Th{1}$ \cr
threshold[t] with error $\frac 1n$	& $\Om{\log \log n}$	& $\Th{1}$ \cr
\hline
\end{tabular}
}\]

\subsection{Relations of quantum circuit classes}

We have shown that $\Bc \QNCwf^0 = \Bc \QACwf^0 = \Bc \QACC^0 = \Bc
\QTCwf^0$ (Theorem~\ref{th:exactt-1}).  If we allow arbitrary one-qubit
gates, then also $\QTCwf^0 = \QACwf^0 \subseteq \QNCwf(\logstar n)$
(Theorems~\ref{th:exactt-ac1} and~\ref{th:exactt-logstar}).
Several open problems of~\cite{ghmp:qacc} have thus been solved.
Only little is known about classes that do not include the fan-out gate.  For
example, we do not know whether $\TC^0 \subseteq \QTC^0$, we only know
that $\TC^0 \subseteq \QTCwf^0$.  It is simple to prove that parity is
in $\TC^0$.  Take the logical Or of exact[1], exact[3], exact[5], \dots, and compute
exact[$k$] from threshold[$k$] and threshold[$k+1$].  However, this
method needs fan-out to copy the input bits and hence it is not in
$\QTC^0$.

Fang et al. proved~\cite{ffghz:fanout} a lower bound for
fan-out.  In particular, they showed that logarithmic depth is needed to
approximate parity using only a constant number of ancillas.
Unfortunately, their method breaks down with more than a linear number
of ancillas and it cannot be extended to other unbounded fan-in gates
such as majority or threshold[t].

\subsection{Upper bounds for $\mbox{\bf B-} \QNCwf^0$}

Shor's original factoring algorithm~\cite{shor:factoring} uses modular exponentiation and
the quantum Fourier transform modulo $2^n$ followed by a polynomial-time deterministic
algorithm.  The modular exponentiation $a^x$ can be replaced by
multiplication of some subset of numbers $a$, $a^2$, $a^4$, \dots,
$a^{2^{n-1}}$~\cite{cw:qft}.  The $n$ numbers $a^{2^k}$ can be quickly
precomputed classically.

Since both multiplication of $n$ numbers (Theorem~\ref{th:arithmetic})
and the QFT (Theorem~\ref{th:qft}) are in $\Bc \QNCwf^0$, there is a
polynomial-time bounded-error classical algorithm with oracle $\Bc \QNCwf^0$ factoring
numbers, i.e.\ factoring $\in \RP [\Bc \QNCwf^0]$.  If $\Bc \QNCwf^0 \subseteq
\BPP$,%
\footnote{In this context, $\Bc \QNCwf^0$ denotes the set of languages
decided with bounded error by constant-depth quantum circuits with
fan-out.}
then factoring $\in \RP [\BPP] \subseteq \BPP [\BPP] = \BPP$.
Discrete logarithms can be computed in a similar way using modular
exponentiation and the quantum Fourier transform modulo general
$q$~\cite{shor:factoring}.  Since QFT$_q \in \Bc \QNCwf^0$
(Theorem~\ref{th:qftq}), we conclude that also discrete-log $\in \RP
[\Bc \QNCwf^0]$.

\subsection{Open problems}

We propose the following open problems on computational aspects of multi-qubit
gates:

\begin{enumerate}
\item Is there a constant-depth exact circuit for Or?

\item Is there a constant-depth linear-size circuit for Or?

\item Are there exact circuits with a fixed basis?

\item Can we simulate unbounded fan-out in constant depth using unbounded
fan-in gates, e.g.\ threshold[t] or exact[t]?
\end{enumerate}

\section*{Acknowledgements}

We thank Harry Buhrman, Hartmut Klauck, and Hein R{\" o}hrig
at CWI in Amsterdam, and Fred Green at Clark University in Worcester
for plenty of helpful discussions, and Ronald de Wolf at CWI for help
with writing the paper.  We thank Klaus M\"olmer and Brian King for
discussions on physical implementations of multi-qubit gates.
We are grateful to Schloss Dagstuhl, Germany, for providing an excellent
environment, where part of this work was carried out.
We thank the anonymous reviewers for their valuable comments.

\nocite{spalek:thesis-qncwf-uploaded}
\nocite{spalek:qncwf-stacs}
\bibliographystyle{alpha}
\bibliography{../quantum}

\appendix

\section{Lower bounds on classical circuits}
\label{app:lower-bounds}

Using the polynomial method~\cite{beigel:polmethod}, we prove several
lower bounds on the depths of deterministic circuits.  We consider
circuits with fan-in of Or and And gates at most 2, and unbounded fan-out
and parity, the same as in the quantum model.

Basically, the value of each bit computed by a circuit can be computed
by a multi-linear polynomial (over the field $\Z_2$) in the input bits.  We
are interested in the degree of such a polynomial; by proving a
lower bound on the degree, we also lower-bound the depth of the circuit.
It is simple to prove that the polynomial computing a Boolean function
is unique.

Each input bit $x_k \in \{ 0, 1 \}$ is computed by the polynomial $x_k$
of degree 1.  The Not gate computes the polynomial $1-p(x)$, where
$p(x)$ is the polynomial computing its argument, and the degree is
unchanged.  The And gate computes the polynomial $p_1(x) \cdot p_2(x)$
and the two degrees are summed.  The parity gate computes the polynomial
$(p_1(x) + \dots + p_k(x)) \mod 2$ of degree equal to the maximum degree
among the arguments.

\begin{lemma}
The output of a circuit of depth $d$ has degree at most $2^d$.
\end{lemma}

\begin{proof}
By induction: by adding a new layer, we can at most double the degree when
using the And gate.
\end{proof}

And of $n$ bits is computed by a (unique) polynomial $x_1 x_2 \dots x_n$
of degree $n$.  Hence every circuit computing And has depth at least $\log
n$.  It is simple to prove by contradiction that also Or, threshold[t],
and exact[t] have full degree $n$.  Smolensky has proved a much stronger
result~\cite{smolensky:algmethods}, which implies that also the degree
of mod[q] for $q>2$ is $n$.

\bigskip

Randomised circuits have access to random bits and may produce the
result with a small error.  Some functions are computed in smaller depth
in this model.

\begin{lemma}
Or can be computed with one-sided error $1 \over 2$ by a randomised
circuit of depth 2.  The error can be decreased to $1 \over n$ in
additional depth $\log \log n$.
\end{lemma}

\begin{proof}
Take $n$ random bits and output the parity $x_1 r_1 \oplus x_2 r_2
\oplus \cdots \oplus x_n r_n$.  If $|x|=0$, then the circuit always
outputs 0.  If $|x|>0$, then the probability that the parity is odd is
equal to $1 \over 2$.
If we perform the computation $(\log n)$-times using independent random
bits, we decrease the probability of error to $({1\over2})^{\log n} = {1
\over n}$.  This can be done in additional depth $\log \log n$ by a
balanced binary tree of Or gates.
\end{proof}

By Yao's principle~\cite{yao:unified}, if we have a randomised circuit with
error less than $2^{-n}$, then there exists an assignment of random bits such that
the result is always correct.  That is there exists a deterministic
circuit of the same shape.  Hence also randomised circuits computing the
logical Or
with exponentially small error have depth at least $\log n$.

\begin{lemma}
Every circuit computing Or with error $1 \over n$ has depth at least $\log
\log n$.
\end{lemma}

\begin{proof}
Assume the converse: there exists a circuit of depth $d < \log \log n$
with error $1 \over n$.  By computing the logical Or independently $n \over \log
n$-times, we can reduce the error to $({1 \over n})^{n \over \log n} =
2^{-n}$.  This can be done in additional depth $\log {n \over \log n} =
\log n - \log \log n$.  The total depth of this circuit is $\log n -
\log \log n + d < \log n$.  However, by Yao's principle, the depth
has to be at least $\log n$.
\end{proof}

\end{document}